# Multi-Physics Mathematical Model of Weakly-Ionized Plasma Flows

**Osama Ahmed Marzouk**

College of Engineering, University of Buraimi, Al Buraimi, Sultanate of Oman

**Email address:**
osama.m@uob.edu.om

**Abstract:** This work presents a multidisciplinary mathematical model, as a set of coupled governing equations and auxiliary relations describing the fluid-flow, thermal, and electric fields of partially-ionized plasma with low magnetic Reynolds numbers. The model is generic enough to handle three-dimensionality, Hall effect, compressibility, and variability of fluid, thermal, and electric properties of the plasma. The model can be of interest to computational modelers aiming to build a solver that quantitatively assesses direct extraction of electric energy from a plasma flow. Three different approaches are proposed to solve numerically for the electric fields with different levels of tolerance toward possible numerical instability encountered at a large Hall parameter, where the effective conductivity tensor loses diagonal dominance and becomes close to singular. A submodel for calculating the local electric properties of the plasma is presented in detail and is applied to demonstrate the effect of different factors on the electric conductivity, including the fuel's carbon/hydrogen ratio and the alkaline seed element that acts as the ionizing species. An analytical expression for the collision cross-section for argon is developed, such that this noble gas can be included as one of the gaseous species comprising the plasma.

**Keywords:** Plasma, Modeling, Hall Effect, Magnetohydrodynamic, MHD Generator

## 1. Introduction

Ionization is an endothermic process in which a sufficient amount of energy exceeding the ionization energy of an atom or molecule is supplied to it, leading to the liberation of an electron. First ionization refers to the removal of the first electron. The removal of the second electron is called second ionization. We here consider only first ionization because it requires the least amount of energy and it is much more encountered in practical and engineering applications.

Ionization is classified into 3 main types based on the source of the supplied energy to overcome the ionization energy. The first method is electron ionization, where the gas is bombarded with an energetic electron beam, stripping off other electrons from the atoms or molecules upon collision. This type is common in mass spectrometry [1]. The second type is ionizing radiation, where sufficiently-powerful phonons (e.g., in the ultraviolet or X-ray radiation spectrum) overcome the binding energy of electrons in atoms or molecules and free them. This type finds uses in the medical field through diagnostic and therapeutic applications [2]. The third type is thermal ionization, which is due to the collision of thermally-agitated atoms or molecules, constituting a hot gas. When these atoms or molecules are very hot, their random-motion velocity becomes high enough so that the kinetic energy transferred in a collision between two atoms or molecules is sufficient to ionize one of them. This type occurs in electric arcs, high-temperature flames, and behind strong shock waves in hypersonic flow fields [3]. We here consider thermal ionization, which is relevant to the application we are concerned with, namely extracting electric energy from weakly-ionized hot-gas plasma in a magnetohydrodynamic (MHD) generator channel. This plasma is formed when a small amount (e.g., 1% by mole) of an alkali metal vapor, especially cesium or potassium [4], is present in the combustion products. This alkali metal can be introduced into the combustor as an alkaline salt compound added to the reactants. For example, potassium carbonate ($K_2CO_3$) can be used for potassium [5]. Alkali metals have low ionization energies, making them much more ionizable than conventional combustion gases (namely $CO_2$ and $H_2O$), which in turn makes them practically the only source of electrons in the hot-gas plasma.

MHD generators are thermoelectric devices that convert chemical energy in the form of a fuel into electricity. The fuel is burned in a combustor using a suitable oxidizer and seeded alkali metal compound. The electric conductivity $\sigma$ of alkali-seeded slightly-ionized plasma is strongly dependent on the absolute temperature $T$ [4], taking roughly the following formula [6], Eq. (1):

$$\sigma = \lambda_1 \exp\left(-\frac{\lambda_2}{T}\right) \qquad (1)$$

where $\lambda_1$ and $\lambda_2$ are coefficients that depend on the plasma gas composition. Thus, a high temperature is critical to the success of the MHD concept. This favors the use of oxygen as the oxidizer rather than air to avoid the dilution effect of nitrogen or other recirculated flue gases (RFG) [7]. The plasma then flows downstream the channel which is subject to an externally applied lateral magnetic field of magnitude ($B$) and has a top electrode (anode) with low voltage and a bottom electrode (cathode) with a high voltage. These voltages can be controlled. For example, the low voltage is zero if the anode is grounded; then the high voltage will be at its largest value if the external electric circuitry is open, but it will be at its smallest value if the external electric circuitry is shorted. Optionally, the combustor exit is fitted with a convergent nozzle ending with a throat upstream of the channel. This makes the flow in the channel supersonic, which critically intensifies the electric output from the channel. In that case, a diffuser (convergent passage) is fitted downstream of the channel to reduce the high velocity of plasma. Figure 1. elucidates the aforementioned components of an MHD generator.

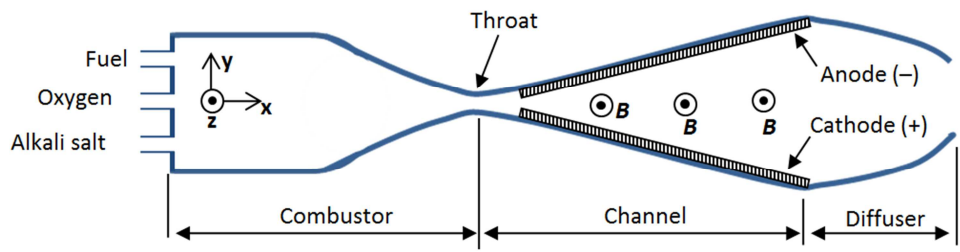

***Figure 1.*** *Sketch of MHD generator.*

The interest in MHD electric power is not new, and it goes back to the early 1970s, where the former USSR has built different MHD generator units as portable, short-term (pulse mode), high-output electric sources for area and deep electrical geophysical surveying [8, 9]. Later, the USA has launched an experimental proof-of-concept (POC) [10] study to explore the technical and economical aspects of coal-fired MHD power generation for retrofitting a plant of 300 MWth (thermal input power). The program ran during 1987 to 1993. Aside from the experimental studies, computational simulations have been performed to predict the performance of potential designs for MHD channels. These studies were contemporary with that phase of interest in MHD generators, and thus were based on simplified mathematical modeling (e.g., steady problem with no time variations [11], axisymmetric [12], one dimensional [13], two-dimensional [14], or three-dimensional but with two-dimensional electric fields and parabolized form of Navier-Stokes equations [15] which eliminates the second-order derivatives in order to allow marching in the axial direction in an technique similar to the boundary-layer techniques [16]) of the plasma motion and electric fields. Recent interest in MHD generation has aroused due to different factors, including stronger magnets which boost the output power from MHD channels, enhanced combustor designs that help lengthening the electrode life, and the advance in computational fluid dynamics (CFD) enabling more reliable prediction of the performance of MHD units [17]. It is the aim of this work to provide a complete mathematical model that describes the temporal and spatial evolution of plasma-gas fields and electric fields within the channel, such that computational researchers can pursue further and use this model in numerical simulations (e.g., based on finite volume discretization) to predict the performance of arbitrary MHD channel designs. The model consists of 3 conservation equations plus the equation of state for the 4 flow fields (density, pressure, stagnation enthalpy, and the velocity vector), and 3 equations for the electric fields (electric potential, current density vector, and electrostatic field vector). Thus, there is a total of 13 unknown scalar fields described by 13 equations. The model is augmented with a submodel to calculate the plasma electric properties, such as the electric conductivity. We used this submodel to analyze multiple factors affecting the electric conductivity. Three different computational approaches are discussed to resolve the electric fields, coping with poor numerical behavior at plasma conditions characterized by a large Hall parameter.

## 2. Elementary Electromagnetic Equations

Before 1864, the laws of electromagnetism were stated in integral forms until James Maxwell reformulated them in differential forms. Maxwell equations constitute of Gauss' law of electricity (Gauss' law), Gauss' law of magnetism (solenoidal law of the magnetic field), Faraday's law of induction, and Ampère-Maxwell equation. Gauss' law, the solenoidal law of a magnetic field, and Ampère-Maxwell equation are not very instrumental in our work. In the following, we present the remaining law among Maxwell equations, namely Faraday's law, plus the generalized Ohm's law, and combine both with the charge conservation law to devise the equations governing the electric fields in an MHD generator channel. This will be manipulated further in section 0.

### 2.1. Faraday's Law

Verbally, Faraday's law mentions that if a conductor (plasma in our case) moves within a magnetic field, where a closed loop $c$ is moving with the plasma, an electromotive force will be generated that is equal to minus the rate of change of the magnetic flux through the loop. In an integral form, this law has the form in Eq. (2). The differential form follows from applying Stokes theorem and is given in Eq. (3). Equation (2) implies that an emf is induced if the applied magnetic field is time-dependent. It should be noted that if the magnetic field is not time-dependent but the conductor is moving, then an emf is still induced in the closed path because there is still a change in the magnetic flux crossing perpendicularly the area spanned by the loop.

$$\text{emf} = \oint_c \vec{E} \cdot d\vec{l} = -\frac{\partial}{\partial t} \iint_S \vec{B} \cdot d\vec{S} \quad (2)$$

$$\nabla \times \vec{E} = -\frac{\partial}{\partial t}\vec{B} \quad (3)$$

The electric field $\vec{E}$ in Eq. (3) can be decomposed into two fields, namely the electrostatic field $\vec{E_s}$ and the unsteady electric field $\vec{E_u}$ induced due to any unsteadiness of the magnetic field [18]. The first electric field is irrotational, by Coulomb's law. Only the second electric is tied to the time rate of change in the magnetic field. Because the magnetic field is not changing in the MHD channel, we are only concerned about the electrostatic field, and we have:

$$\nabla \times \vec{E_s} = 0 \quad (4)$$

The above equation means that an electric potential function $\phi$ can be introduced such that

$$\vec{E_s} = -\nabla \phi \quad (5)$$

There is another electric field induced due to the motion of the plasma (the conductor) with velocity $\vec{u}$ within the applied magnetic field $\vec{B}$. This motional electric field is equal to $\vec{E_m} = \vec{u} \times \vec{B}$. This can be shown from Eq. (2), and it is the working principle for the turbo-generators. When applying Eq. (3) for that case, one should note that the change in the magnetic field is with respect to a moving observer traveling with the plasma, thus the closed loop in Eq. (2) is moving rather than being stationary. Moreover, the drift of the electrons (as a conductor) within the plasma also induces another electric field, manifesting the Hall effect. So, in general we have

$$\vec{E} = \vec{E_s} + \vec{E_u} + \vec{u} \times \vec{B} + \vec{v_d} \times \vec{B} \quad (6)$$

The second term is zero for our applications because we consider only the case a time-fixed (steady) magnetic field. Theoretically, a magnetic field ($\vec{b}$) due to the currents within the plasma is induced but it is safely neglected here. This is a very reasonable assumption for an MHD channel, called low magnetic Reynolds number, and is described in Appendix A.

### 2.2. Generalized Ohm's Law

The (simple) Ohm's law for a conducting medium subject to an electrostatic field simply establishes a linear relation between the current density within the medium and that electrostatic field, thus

$$\vec{J} = \sigma \vec{E_s} \quad (7)$$

It is to be noted that the current density vector is collinear with the electrostatic field and has the same direction. However, in the presence of other induced electric fields as a result of any unsteadiness of the magnetic field, the motion of the plasma under the effect of an applied magnetic field, or the drift of the electrons within the plasma under the effect of an applied magnetic field, the law is generalized as:

$$\vec{J} = \sigma\left(\vec{E_s} + \vec{u} \times \vec{B}\right) - \mu_e\left(\vec{J} \times \vec{B}\right) \quad (8)$$

The three terms in the RHS represent electrostatic current density, Faraday current density, and Hall current density, respectively. The derivation of Eq. (8) is given in Appendix B.

### 2.3. Charge Conservation Law

Because the charge is a conserved quantity, the net flux of $\vec{J}$ leaving out of an arbitrary closed surface $S$ enclosing a volume $\Psi$ is equal to the time rate of decrease of the charge contained in that volume. In its integral form, the charge conservation law is expressed as

$$\iint_S \vec{J} \cdot d\vec{S} = -\frac{\partial}{\partial t} \iiint_\Psi \rho_{ch} \, d\Psi \quad (9)$$

Using Gauss' theorem, the differential form of this law is

$$\nabla \cdot \vec{J} = -\frac{\partial \rho_{ch}}{\partial t} \quad (10)$$

For our purposes, the RHS of Eq. (10) is zero since the plasma is macroscopically neutral (in general, this term is negligible in conductors [18]), and we have

$$\nabla \cdot \vec{J} = 0 \quad (11)$$

## 3. Derived Electromagnetic Equations

In this section, we manipulate the elementary electromagnetic equations presented in the previous section to derive partial differential equations that are more suitable for implementation computationally, by augmenting them to an arbitrary fluid-dynamic solver.

### 3.1. Generalized Ohm's Law with Tensorial Conductivity

The generalized Ohm's law, Eq. (8), can be expressed in a more compact form that is explicit in the current density (i.e.,

$\vec{J}$ appears only in the LHS). To achieve this, we recall that the vector product $(\vec{J} \times \vec{B})$ can be expressed as the inner product of a skew-symmetric tensor $\vec{\vec{B}}$ and $\vec{J}$ (as a first-rank tensor). The tensor $\vec{\vec{B}}$ has the following form:

$$\vec{\vec{B}} = \begin{bmatrix} 0 & B_z & -B_y \\ -B_z & 0 & B_x \\ B_y & -B_x & 0 \end{bmatrix} \quad (12)$$

where $B_x, B_y, B_z$ are the Cartesian components of the magnetic field vector $\vec{B}$. Using this vector rule into the generalized Ohm's law, Eq. (8), and slightly manipulating the terms, we obtain

$$\left[\vec{\vec{I}} + \mu_e \vec{\vec{B}}\right] \vec{J} = \left[\sigma \vec{\vec{I}}\right] \cdot (\vec{E_s} + \vec{u} \times \vec{B}) \quad (13)$$

where $\vec{\vec{I}}$ is the identity tensor. Denoting the coefficient tensor in the LHS as $\vec{\vec{A}}$, and multiplying the above equation by the inverse of this tensor, we obtain

$$\vec{J} = \left[\sigma \overrightarrow{\overrightarrow{A^{-1}}}\right] \cdot (\vec{E_s} + \vec{u} \times \vec{B}) \quad (14)$$

or

$$\vec{J} = \overrightarrow{\overrightarrow{\sigma_{eff}}} \cdot (\vec{E_s} + \vec{u} \times \vec{B}) \quad (15)$$

where the tensor $\overrightarrow{\overrightarrow{\sigma_{eff}}}$ represents an effective electric conductivity, which is calculated as

$$\overrightarrow{\overrightarrow{\sigma_{eff}}} = \sigma \left[\vec{\vec{I}} + \mu_e \vec{\vec{B}}\right]^{-1} = \sigma \overrightarrow{\overrightarrow{A^{-1}}} \quad (16)$$

Using the symbolic manipulation package SymPy [19] (a Python [20] library for symbolic mathematics), we determined the analytic expression of $\overrightarrow{\overrightarrow{\sigma_{eff}}}$, and this is given in Appendix C.

### 3.2. Poisson Equation for Electric Potential $\phi$

Using the definition of the electric potential, Eq. (5), into the modified generalized Ohm's law, Eq. (15), and then applying the divergence operator to both sides, and recalling that the divergence of the LHS vanishes by the charge conservation law, Eq. (11), we obtain an elliptic Poisson-type partial differential equation for the electric potential $\phi$

$$\nabla \cdot \left(\overrightarrow{\overrightarrow{\sigma_{eff}}} \cdot \nabla \phi\right) = \nabla \cdot \left(\overrightarrow{\overrightarrow{\sigma_{eff}}} \cdot (\vec{u} \times \vec{B})\right) \quad (17)$$

## 4. Calculating the Electric Properties

In order to handle the equations described in the previous section, it is apparent that we need a submodel to calculate the local electric conductivity $\sigma$ and the electron mobility $\mu_e$ as a function of other local plasma parameters, such as the temperature and chemical composition. We propose a submodel based on the equilibrium ionization model of Saha [21, 22], which predicts the number density of free electrons in the plasma at a specified equilibrium temperature. Saha equation for alkali metals gives that, under equilibrium ionization, we have the following expression for the equilibrium constant $\lambda$

$$\lambda \equiv \frac{n_e^2}{n_{s,fin}} = C_{Saha} \, T^{1.5} \exp\left(-\frac{IE'}{k_B' \, T}\right)$$

with $C_{Saha} = \left(\frac{2\pi \, m_e \, k_B}{h^2}\right)^{1.5}$ (18)

where $n_e$ is the number density of the free electrons after ionization (per m³), $n_{s,fin}$ is the number density of seed alkaline atoms after ionization (per m³), and $C_{Saha}$ is a lumped constant which has the value of $2.4247 \times 10^{21}$ (in m⁻³·K⁻¹·⁵). The ionization energy for cesium (Cs) and potassium (K) are 3.8939 eV and 4.3407 eV, respectively [23].

From the conservation of alkaline nuclei, we have

$$n_{s,ini} = n_{s,fin} + n_e \quad (19)$$

where $n_{s,ini}$ is the number density of seed alkaline atoms before ionization (per m³), which is calculated as [24]

$$n_{s,ini} = \frac{X_{s,ini} \, p_{ini}}{k_B T} \quad (20)$$

where $X_{s,ini}$ is the mole fraction of the alkaline atoms before ionization (dimensionless), $p_{ini}$ is the total (not partial) pressure of the gases before ionization (in Pa = N/m²).

In the Eq. (19), the number density of electrons $n_e$ is used instead of the number density if alkaline ions in the RHS, because both have the same values under single-ionization. Combining Eq. (18) and Eq. (19), we obtain a quadratic equation for the electron number density after ionization $n_e$, whose only acceptable (positive) solution is

$$n_e = \frac{-\lambda + \sqrt{\lambda^2 + 4\lambda \, n_{s,ini}}}{2} \quad (21)$$

Once this electron number density is known, the composition of the plasma is known. One can then use the collision cross sections of the neutral species in the plasma along with the number density of each of them and obtain the mean collision frequency $\nu_0$ due to these gaseous species according to Eq. (22) [25]

$$\nu_0 = c_{e,mean} \sum_N n_H Q_{eN} \quad (22)$$

with $c_{e,mean} = \sqrt{\frac{8 k_B T}{\pi m_e}}$

or

$$c_{e,mean} = C_{ce} \sqrt{T}, \text{ where } C_{ce} \equiv \sqrt{\frac{8 k_B}{\pi m_e}} \quad (23)$$

where $c_e$ is the mean of the magnitude of electron velocity based on Maxwell speed distribution (in m/s), $n_N$ is the number density (in 1/m³) of the species $N$, $Q_{eN}$ is the average momentum-transfer cross section for electron collisions with the molecules of the species $N$ (in m$^2$), $k_B$ is Boltzmann constant, $T$ is the absolute temperature (in kelvins). The lumped constant $C_{ce}$ has the value $6212\,\frac{m/s}{\sqrt{K}}$. It should be noted that Eq. (22) is suitable for slightly-ionized plasma rather than fully-ionized plasma, as it does not account for electron-ion and electron-electron collisions.

We used analytical fits (based on experimental data) for the collision cross sections, as a function of the electron energy in eV (electron volt) made by Frost [26]. For Argon, we deduced a fitting function from the data in Ref. [27], because argon is not covered in the work of Frost. We present the details of these analytical fits in Appendix D.

The mean collision frequency due to neutrals only is augmented by the mean collision frequency due to ions (and electrons) scattering $\nu_1$ (in Hz), to yield the total mean collision frequency $\nu_{tot}$.

$$\nu_{tot} = \nu_0 + \nu_1 \qquad (24)$$

where

$$\nu_1 = 0.476\,\frac{\kappa}{\eta}$$

where

$$\kappa = C_\kappa\, n_e\, \frac{\ln(\Lambda)}{(k'_B T)^{1.5}}$$

$$C_\kappa = \frac{2\pi e^2}{(4\pi \varepsilon_0)^2}\sqrt{2\,\frac{e}{m_e}}$$

$$\Lambda = \frac{C_\Lambda (k'_B T)^{1.5}}{\sqrt{n_e}}$$

$$C_\Lambda = \frac{3}{\sqrt{4\pi}}\left(\frac{4\pi \varepsilon_0}{e}\right)^{1.5} \qquad (25)$$

In the above set of equations, $\varepsilon_0$ is the electric permittivity of vacuum ($8.854 \times 10^{-12}$ F/m), the lumped constant $C_\kappa$ has the value $7.727 \times 10^{-12}$ (when using SI units), and the lumped constant $C_\Lambda$ has the value $1.549 \times 10^{13}$ (when using SI units). Once the $\nu_{tot}$ frequency is calculated, we proceed with calculating the electron mobility as described later in the Nomenclature section, in the part devoted there to the electron mobility, but with $\nu_{tot}$ replacing $\nu_e$.

Comparisons with reference data [6] support the validity of our implementation of this submodel, which we carried out using the Python platform. Another comparison is made with a case of air at 3273 K with 2% potassium [28]. The cited value of the magnetic Reynolds number (Re$_m$) for this case is $1.3\times10^{-5}$, under a reference velocity of 1 m/s and a reference length of 1 m. Taking the magnetic permeability of air to be the same as the one of vacuum, i.e., $4\pi \times 10^{-7}$ H/m, (because there are very close to each other), and from the definition of Re$_m$ in Eq. (A-1), we infer that the corresponding electric conductivity for this case is 103 S/m. Our calculation estimates the electric conductivity for this case to be 192 S/m. Although the difference is not small, the agreement is thought to be favorable in the light of large uncertainty in the analytical fits, and the lack of details about the limitations and calculation method in the source reference. Relatively large deviations in $\sigma$ across different sources are acceptably waived [4]. In our calculations, we approximated the air as a mixture of 21% O$_2$ and 79% N$_2$, by mole. We also assumed that the pressure is atmospheric; the pressure value in the source reference is not disclosed. In fact, if the pressure in the reference source is high, then our calculation will approach the cited value. For example, if the pressure is twice the atmospheric value, our predicted electric conductivity drops to 158 S/m, and drops further to 139 S/m if the pressure is increased by another atm. We performed another validation against an experimental value of $\sigma$ [29] for combustion-plasma of a paraffin fuel (kerosene) having a C/H ratio of 0.5 (1 carbon atom per 2 hydrogen atoms [30]) which is typical for transportation fuels [31], in oxygen at two temperatures with 1% potassium by mole. We also compared our $\sigma$ to a computational value [4]. The comparison is summarized in Table 1, and our results fairly agree with the measurements.

*Table 1.* Comparison of electric conductivity with other sources for a test-case.

| Temperature [K] | Measured $\sigma$ [S/m] | Cited calculation | Our calculation |
|---|---|---|---|
| 2800 | 25 | 20 | 30.8 |
| 3000 | 65 | 60 | 57.9 |

The submodel should be applied at each spatial point where the electric properties of the plasma are sought. The inputs to the submodel include the local volume fraction of the alkali metal vapor (either cesium or potassium) and the mole fractions of potential gaseous molecules which may appear in the combustion products. Due to the relatively high ionization energies of all species except the alkali metal, the free electrons are taken to be due to the ionization of the alkali metal only. However, the other species affect the electron mobility because they form inhibit a collisionless electron motion to different degrees, depending on their momentum cross sections. This submodel fits well the plasma conditions in an MHD channel, where the ionization is limited to a small fraction (≈1%) of the alkaline vapor, which is a small fraction (≈1%) of the plasma gas.

## 5. Results and Discussion

This section is divided into three main parts; the first and second parts are devoted to presenting the equations governing the thermo-fluid and electric field variables of the plasma, with proposed algorithms to solve for the electric fields. In the last part, we shed light on important factors that affect the electric conductivity of the plasma, through applying the electric submodel described in the previous section.

### *5.1. Governing Equations – Fluid-Flow Part*

In this section, we put the partial differential equations describing the evolution of the velocity vector, temperature,

pressure, and density of the plasma-gas [32], with suitable source terms to couple the electric fields to the fluid-flow fields.

There is no major restricting assumption about the fluid regime or dimensionality, and a generic compressible three-dimensional flow with non-constant momentum and thermal properties is treated. Similarly, the governing equations for the electric fields do not stipulate a restriction on the dimensionality of these fields or a predefined spatial distribution for the electric conductivity or electron mobility.

### 5.1.1. Mass Conservation

The mass conservation equation of the plasma gas is not affected by the presence of electric fields. The equation preserves its differential form for a non-conducting fluid, being

$$\frac{\partial \rho}{\partial t} + \nabla \cdot (\rho \vec{u}) = 0 \qquad (26)$$

where $\rho$ is the mass density of the plasma gas.

### 5.1.2. Momentum Conservation

The momentum conservation equation of the plasma gas derives from the original vectorial Navier-Stokes equation, but a source term is added representing the force (per unit volume) acting on the charged particles. Any translating conductor carrying a current while subject to a magnetic field will experience a force called the Lorentz force [18], perpendicular to both the motion and the magnetic field. Its origin comes from the force acting on an individual charged particle with a charge $q$ (in coulombs), where we have

$$\vec{F_q} = q \, \vec{u} \times \vec{B} \qquad (27)$$

where $q = -e$ for an electron, but $q = e$ for an ion. Applying this equation to the electrons and ions within a unit volume of the plasma, we obtain

$$\vec{F_{V=1}} = -e \, n_e \, \vec{u_e} \times \vec{B} + e \, n_e \, \vec{u_{ion}} \times \vec{B} \qquad (28)$$

Given the large ion-to-electron mass ratio, the ion is considered to move with the same bulk velocity of the plasma-gas, thus $\vec{u_{ion}} = \vec{u}$. This allows us to re-write the above equation as

$$\vec{F_{V=1}} = -e \, n_e \, (\vec{u_e} - \vec{u}) \times \vec{B} = -e \, n_e \, (\vec{v_d}) \times \vec{B} \qquad (29)$$

But (from the Nomenclature section provided later, under the part devoted there to the electric-current density, $\vec{J}$) this means

$$\vec{F_{V=1}} = \vec{J} \times \vec{B} \qquad (30)$$

This is the Lorentz force per unit volume, which is a body force exerted on a unit volume of the plasma in the MHD channel due to the presence of electromagnetic fields. It is the means of coupling the electric field into the motion of the plasma. The modified momentum equation with the additional RHS source term is thus

$$\frac{\partial \rho \vec{u}}{\partial t} + \nabla \cdot (\rho \, \vec{u} \otimes \vec{u}) + \nabla p - \nabla \cdot \vec{\vec{\tau}} = \vec{J} \times \vec{B} \qquad (31)$$

where $p$ is the pressure of the plasma gas, and $\vec{\vec{\tau}}$ is the viscous stress tensor.

We point out that in the limiting case of one-dimensional plasma velocity and magnetic field and electrostatic field (mutually orthogonal to each other), the Lorentz source term reduces to $-|J|B$. The minus sign of this term reveals the dissipative nature of this source, which is in fact a sink of plasma momentum, trying to decelerate the plasma and/or decrease its pressure as it progresses downstream the MHD channel.

As for any Newtonian fluid, the viscous stress tensor $\vec{\vec{\tau}}$ is modeled using the following constitutive relation [33]

$$\vec{\vec{\tau}} = \mu \left( \overline{\overline{\nabla u}} + \left[ \overline{\overline{\nabla u}} \right]^T - \frac{2}{3} (\nabla \cdot \vec{u}) \, \vec{\vec{I}} \right) \qquad (32)$$

where $\mu$ is the dynamic viscosity of the plasma-gas, $\overline{\overline{\nabla u}}$ is the velocity gradient tensor, $\left[ \overline{\overline{\nabla u}} \right]^T$ is its transpose, and $\vec{\vec{I}}$ is the identity tensor.

### 5.1.3. Energy Conservation

There are several ways to express the energy conservation of a fluid. We here use one form convenient to high-speed gases, which is demanded for MHD channels. This form utilizes the stagnation specific enthalpy $H$ as the dependent variable rather than the temperature. A nonlinear solver is needed to extract the temperature from $H$, where we first compute the static specific enthalpy $h$ as

$$h = H - \frac{1}{2} \vec{u} \cdot \vec{u} \qquad (33)$$

and then solve the following nonlinear equation for the temperature

$$h = h_{Ref}(T_{Ref}) + \int_{T_{Ref}}^{T} C_p(T) \, dT \qquad (34)$$

The specific heat at constant pressure $C_p(T)$ is customarily expressed as a piecewise polynomial of temperature [34], with a reference specific enthalpy $h_{Ref}$ at a reference temperature $T_{Ref}$ for each species.

The partial differential equation for the energy conservation is

$$\frac{\partial \rho H}{\partial t} + \nabla \cdot (\rho \, \vec{u} \, H) - \frac{\partial p}{\partial t} + \nabla \cdot \vec{q} + \nabla \cdot (\vec{u} \cdot \vec{\vec{\tau}}) = \vec{J} \cdot \vec{E_s} \qquad (35)$$

where $\vec{q}$ is the diffusion heat flux vector, which is typically expressed using the following constitutive relation (Fourier's law in this case) [35]:

$$\vec{q} = -\kappa \, \nabla T \qquad (36)$$

where $\kappa$ is the thermal conductivity (in W/m·K).

The MHD scalar source term in Eq. (35) expresses the power (per unit volume) extracted from the plasma and converted into electric power to an external load connected to the channel electrodes. This term is negative, acting actually as an energy sink (not source), as expected. In the limiting case of one-dimensional plasma velocity and magnetic field and electrostatic field (mutually orthogonal to each other), this source term reduces to $-|J|E$.

*5.1.4. Ideal Gas Law*

With $R$ being the specific gas constant (in J/kg·K), the relation between the pressure, density, and temperature follows the following form of the ideal gas law

$$p = \rho R T \qquad (37)$$

Because combustion plasma is composed of more than one gaseous species, the $R$ value in the above equation is an average value, calculated by

$$R = \frac{\bar{R}}{\bar{M}} \qquad (38)$$

where $\bar{M}$ is the molecular weight of the mixture, to be calculated as [36]

$$\bar{M} = \sum_i X_i M_i \qquad (39)$$

and $X_i$ and $M_i$ are the mole fraction and molecular weight of each constituent gaseous species in the plasma.

*5.2. Governing Equations – Electric Part*

We have already presented in sections 0 and 0 the elementary equations governing the electric-field variables (namely $\vec{E_s}$, $\vec{J}$, and $\phi$) within the MHD channel. Out of these 3 electric fields, $\vec{E_s}$ and $\vec{J}$ are mandatory because they form the MHD source terms in the fluid-flow momentum and energy equations. We recall that the magnetic field $\vec{B}$ is part of the momentum source term, but its spatial profile should be a given input rather than an unknown to solve for. We propose below 3 different computational approaches to solve for $\vec{E_s}$ and $\vec{J}$; all of which solve for the scalar potential function $\phi$ first.

*5.2.1. Solving the Electric Fields: Approach 1*

This approach is the most-straightforward, where it solves the tensor-based Poisson equation for the potential scalar $\phi$. The equation is repeated below for convenience.

$$\nabla \cdot \left(\overline{\overline{\sigma_{eff}}} \cdot \nabla \phi\right) = \nabla \cdot \left(\overline{\overline{\sigma_{eff}}} \cdot (\vec{u} \times \vec{B})\right)$$

The LHS is discretized implicitly, whereas the source RHS is treated explicitly. When integrating Eq. (17) numerically, we can apply a Dirichlet (or first-type) boundary condition at the anode and cathode, fixing the value of $\phi$ at each electrode to the desired value, for example

$$\phi_{anode} = 0$$

$$\phi_{cathode} = V_{Load} \qquad (40)$$

Thus, $\phi = 0$ at the low-voltage anode (as if it is grounded), but $\phi = V_{Load}$ at the high-voltage cathode, and $V_{Load}$ is the desired voltage difference across the channel which is also the voltage difference across the external load connected to the electrodes (if voltage drops in the external circuitry is neglected). Even if the anode is not grounded, we can still set its potential as zero because the absolute value of $\phi$ is not important, but its spatial variation is what matters.

The boundary condition at the other (electrically non-conducting) channel walls should be a Neumann (or second-type) boundary condition of the form [37]

$$\frac{\partial \phi}{\partial n} = (\vec{u} \times \vec{B})_{bound} \cdot \hat{n} \qquad (41)$$

where $\hat{n}$ is a unit normal pointing perpendicularly away from the boundary. Once the scalar electrical potential function $\phi$ is solved for, a discrete gradient operator is applied to obtain the electrostatic field using Eq. (5),

$$\vec{E_s} = -\nabla \phi$$

and then the modified (with tensorial conductivity) generalized Ohm's law, Eq. (15), is applied to find the spatial distribution of the vector current density $\vec{J}$ at the current time instant.

$$\vec{J} = \overline{\overline{\sigma_{eff}}} \cdot \left(\vec{E_s} + \vec{u} \times \vec{B}\right)$$

*5.2.2. Solving the Electric Fields: Approach 2*

Whereas the 1st approach can work smoothly, the numerical behavior may deteriorate if the electron mobility $\mu_e$ (thus the Hall parameter $\beta$) becomes high, where in this case the $\overline{\overline{\sigma_{eff}}}$ becomes less positive-definite. As an alternate (2nd) approach, the LHS of Eq. (17) is kept unchanged, but the source RHS will be adapted and expressed in terms of $\phi$ and $\vec{J}$ rather than $\vec{u}$ and $\vec{B}$. To this end, the modified (with tensorial conductivity) generalized Ohm's law is manipulated, and re-written as,

$$\overline{\overline{\sigma_{eff}}} \cdot (\vec{u} \times \vec{B}) = \vec{J} - \overline{\overline{\sigma_{eff}}} \cdot \vec{E_s} \qquad (42)$$

Using the above equation into the RHS of the tensor-based Poisson equation, $\nabla \cdot \left(\overline{\overline{\sigma_{eff}}} \cdot (\vec{u} \times \vec{B})\right)$, we can re-write this RHS source as $\nabla \cdot \vec{J} - \nabla \cdot \left(\overline{\overline{\sigma_{eff}}} \cdot \vec{E_s}\right)$. Recalling that $\vec{E_s} = -\nabla\phi$, the RHS can be manipulated further and re-written as $\nabla \cdot \vec{J} + \nabla \cdot \left(\overline{\overline{\sigma_{eff}}} \cdot \nabla\phi\right)$. This leads to the following alternate tensor-based Poisson equation for $\phi$,

$$\nabla \cdot \left(\overline{\overline{\sigma_{eff}}} \cdot \nabla \phi\right) = \nabla \cdot \vec{J} + \nabla \cdot \left(\overline{\overline{\sigma_{eff}}} \cdot \nabla \phi\right) \qquad (43)$$

As in the 1st approach, the LHS is treated implicitly, whereas the 2 RHS terms are treated explicitly. It should be mentioned that while the first RHS term is analytically zero by virtue of Eq. (11), it will evaluate to a non-zero value during the calculations, which partly drives the solution of the elliptic equation. Once the scalar electrical potential function $\phi$ is solved for, $\vec{E_s}$ and $\vec{J}$ are obtained using algebraic equations in the same manner described in the 1st approach, i.e., using Eq. (5) for $\vec{E_s}$ and Eq. (15) for $\vec{J}$.

*5.2.3. Solving the Electric Fields: Approach 3*

In the 1st and 2nd approach, the solution of $\phi$ was based on the tensorial $\overline{\overline{\sigma_{eff}}}$, which exhibits a poor numerical character for implicit discretization as the electron mobility (thus the Hall parameter $\beta$) increases. In such case, $\overrightarrow{\sigma_{eff}}$ loses the diagonal dominance and approaches singularity, with its condition number growing to high levels as can be seen from Eq. (C-5) in Appendix C. This may lead to a failure in the computation even with the remedy of the 2$^{nd}$ approach. A 3$^{rd}$ approach is presented here, which does not use $\overrightarrow{\sigma_{eff}}$ altogether. Instead, the scalar $\sigma$ is used, forming an alternate scalar-based Poisson equation for $\phi$. The derivation of this equation is similar to the one given for Eq. (43), but we start from a simplified version of the original generalized Ohm's law, Eq. (8), where the Hall term is omitted, i.e.,

$$\vec{J} = \sigma \left( \vec{E_s} + \vec{u} \times \vec{B} \right) \quad (44)$$

Taking the divergence and using the analytical condition that $\nabla \cdot \vec{J} = 0$ and the relation $\vec{E_s} = -\nabla \phi$, we obtain a simplified Poisson equation for $\phi$, having the form

$$\nabla \cdot (\sigma \nabla \phi) = \nabla \cdot \left( \sigma \left( \vec{u} \times \vec{B} \right) \right) \quad (45)$$

Going back to the simplified Ohm's law, Eq. (44), we re-write it as

$$\sigma \left( \vec{u} \times \vec{B} \right) = \vec{J} - \sigma \vec{E_s} = \vec{J} + \sigma \nabla \phi \quad (46)$$

Then, we take the divergence of both sides,

$$\nabla \cdot \left( \sigma \left( \vec{u} \times \vec{B} \right) \right) = \nabla \cdot \vec{J} + \nabla \cdot (\sigma \nabla \phi) \quad (47)$$

And we use this expression to replace the RHS source in Eq. (45), and again express this source in terms of $\phi$ and $\vec{J}$ rather than $\vec{u}$ and $\vec{B}$, as done in the 2$^{nd}$ approach. We now obtain the final simplified Poisson equation to be attempted in this 3$^{rd}$ approach

$$\nabla \cdot (\sigma \nabla \tilde{\phi}) = \nabla \cdot \vec{J} + \nabla \cdot (\sigma \nabla \tilde{\phi}) \quad (48)$$

The tilde above $\phi$ is used just to emphasize that solving this equation gives a estimate of the electric potential (thus, denoted by $\tilde{\phi}$), because the Hall effect was not accounted for. The LHS is treated implicitly whereas the RHS is an explicit source. After this step, an estimated electrostatic field is obtained as

$$\vec{E_s} = -\nabla \tilde{\phi} \quad (49)$$

Then, a series of stationary-iterative correction is performed for the current density $\vec{J}$ using this estimated $\vec{E_s}$ through the scalar-based generalized Ohm's law in Eq. (50). This correction recovers the ignored Hall contribution to $\vec{J}$. Thus, the following equation is solved iteratively; in every iteration, an updated (more-corrected) value of $\vec{J}$ in the LHS using the old value of $\vec{J}$ in the RHS.

$$\vec{J} = \sigma \left( \vec{E_s} + \vec{u} \times \vec{B} \right) - \mu_e (\vec{J} \times \vec{B}) \quad (50)$$

After a pre-specified number of corrections (or a tolerance-based termination criterion, testing that $\nabla \cdot \vec{J} = 0$ is satisfied within an acceptable tolerance in the local computational cell), we end up with the final value of $\vec{J}$. While this approach may be the most tolerant among the presented 3 approaches for the electric field, there is a lagged influence of the Hall effect on the electric field $\vec{E_s}$. However, as the problem reaches a steady-state condition, the error fades away.

### 5.3. Analysis of Electric Conductivity

The electrical conductivity is a factor affecting directly the amount of electricity that can be extracted from plasma [38]. Careful understanding of its dependence on plasma conditions is thus important. We implemented the electric submodel presented in section 4 using a Python code, and we here examine quantitatively the effect of the plasma-gas composition on the $\sigma$ for a typical MHD channel condition having a temperature of 2800 K, a potassium volume fraction of 1%, and a total pressure of 3 atm (both being before the ionization). The estimated electric conductivity is compared for different gaseous species in Table 2 (the plasma-gas is composed only of that species for each case). This exposes the retarding effect of the molecules on the mobility of the free electrons, and thus on the electric conductivity. We augment also the fictitious case of zero collision cross section, which corresponds to the limiting value of $\sigma$. A base case of air (79% $N_2$ + 21% $O_2$, by volume) is also added as a reference. The last two columns represent the estimated $\sigma$ if only the electron-neutral collisions are considered (while the scattering by ions and electrons is completely ignored - corresponds to setting $\nu_1 = 0$ in Eq. (24)), and the relative erroneous gain in $\sigma$ due to this simplification. The species are ordered with an ascending order of $\sigma$. Argon is distinct in its low resistance for electron motion, giving a $\sigma$ value very close to the upper bound (the ratio is 0.977). Special attention should be paid to the typical major gases in combustion products, namely $CO_2$, $H_2O$, and $N_2$. $CO_2$ and $N_2$ have a similar collision effect, which is far less than the effect of $H_2O$. With $H_2O$, $\sigma$ is only 1/3 of its value in the case of $CO_2$. This means that the hydrocarbon fuel type has a notable impact on the MHD channel performance, because the larger carbon content, the larger plasma conductivity and the better power generation, and vice versa. The use of low C/H fuel should be avoided. The base case (air) lies midway in the table. For species with larger $\sigma$, the error in the simplified $\sigma$ grows, with a limiting upper bound of 46.4% at this temperature and pressure. However, for the 3 typically major gases in combustion products, the error is limited to 13.1%. This alleviates the effect of the $\sigma$ correction in combustion plasma especially that the uncertainties in the model itself are comparable to such error.

*Table 2. Effect of the plasma-species and ion-and-electron scattering on the electric conductivity.*

| Species | Corrected $\sigma$ [S/m] | % of Base | Simplified $\sigma$ [S/m] | % of Detailed $\sigma$ |
|---|---|---|---|---|
| Q=0 | 101.06 | 261% | 147.97 | 46.4% |
| Ar | 98.77 | 255% | 143.1 | 44.9% |
| Ne | 85.84 | 222% | 117.5 | 36.7% |
| $O_2$ | 59.27 | 153% | 72.81 | 22.8% |
| He | 55.99 | 144% | 67.92 | 21.3% |
| CO | 42.05 | 109% | 48.43 | 15.2% |
| O | 41.93 | 108% | 48.28 | 15.1% |

| Species | Corrected $\sigma$ [S/m] | % of Base | Simplified $\sigma$ [S/m] | % of Detailed $\sigma$ |
|---|---|---|---|---|
| Air (base) | 38.72 | 100% | 44.07 | 13.8% |
| $CO_2$ | 36.85 | 95.2% | 41.66 | 13.1% |
| $N_2$ | 35.45 | 91.6% | 39.88 | 12.5% |
| $H_2$ | 32.50 | 83.9% | 36.18 | 11.3% |
| OH | 15.00 | 38.7% | 15.74 | 4.93% |
| $H_2O$ | 12.50 | 32.3% | 13.01 | 4.08% |
| H | 10.52 | 27.2% | 10.88 | 3.42% |

To further examine the factors affecting $\sigma$, we show in Figure 1 its dependence on the temperature in the range 2200 – 3400 K where the gas is air (same base case in Table 2), with either potassium or cesium used as the ionizing alkali metal. This temperature range covers the reasonable operating temperature for an MHD channel. The lower half of the range is relevant to air-fired MHD combustor, where the presence of large amount of diluting air-born $N_2$ reduces the temperature. On the other hand, the upper half of the temperature range is more relevant to oxygen-fired MHD combustor, where the absence of $N_2$ raises the combustion temperature. It is evident that the latter combustion technique greatly benefits from the large boost in $\sigma$ due to the elevated temperatures. The amplification in $\sigma$ (we refer to the corrected value, by default) as a result of using cesium instead of potassium is clear, and is a direct consequence of the lower ionization energy of the former. This amplification decreases from 3.15 at 3400 K to 1.63 at 2200 K. We also compare in Figure 1 the corrected and simplified $\sigma$ (i.e., with and without the $v_1$ correction, respectively). We make the same note made for the data presented for Table 2: the correction effect is pronounced at large $\sigma$. This can be attributed to the larger amount of free electrons, which intensifies the ion-electron and electron-electron scatterings, which are captured by the correction term, reducing the over-predicted simplified $\sigma$.

We also point out that the electric conductivity is very sensitive to the temperature, but the dependence is not accurately approximated as the exponential function as in Eq. (1), which will under-predict $\sigma$. The presence of the power term in the Saha equation and the nonlinear collision effects complicate the overall dependence, but it is better approximated as a quadratic profile based on our curve-fitting analysis.

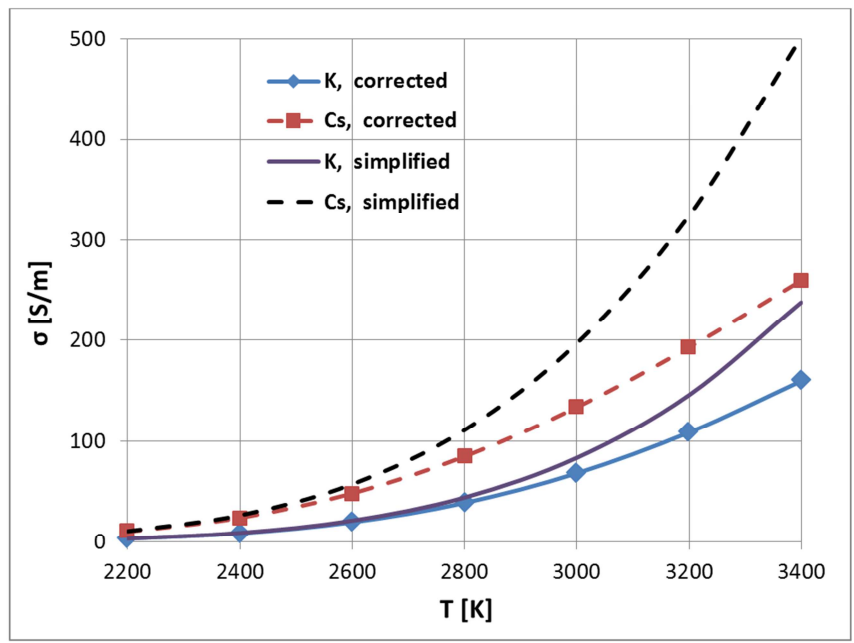

*Figure 1.* *Effect of temperature, seed metal, and scattering correction on the electric conductivity.*

## 6. Conclusion

We presented a set of partial differential equations for describing the temporal and spatial evolution of the three-dimensional fluid-flow variables and electric variables of alkali-seeded combustion plasma subject to an applied magnetic field. Derivations and auxiliary expressions are given and a modified form of the generalized Ohm's law that is explicit in the current density vector was derived. The model is supported with a complete submodel for predicting the electric properties of the plasma locally, accounting for scattering due to neutrals, electrons, and ions. This submodel is a prerequisite for applying the presented mathematical model. The fluid-flow equations can be integrated numerically in space and time using one of the several available computational fluid dynamics (CFD) solvers, utilizing the finite volume or the finite difference method. To make the flow solver capable of handling the Faraday induction and Hall effect for a plasma gas, one should augment one source term into each of the momentum and energy equations, and solve for the electrostatic field and the current density. Different computational algorithms were proposed to solve for the electric fields, all of them solve a Poisson equation for the electric scalar potential, but the formulation and solution procedure vary to allow coping with computational instabilities at high electron motilities and Hall parameters; in which case the conductivity tensor of the Poisson equation becomes nearly skew-symmetric. The sensitivity of the electric conductivity, a key parameter in model, to different design choices was discussed in light of the implemented electric-properties submodel. Fuels which are richer in carbon than hydrogen lead to appreciably elevated electric conductivity, thus more electricity extraction. Correction due to electron-and-ion scattering cannot be ignored for conductivity values approximately above 40 S/m.

The mathematical model is a preliminary step toward building a computational electro-fluid dynamics (CEFD) solver to act as a tool for predicting the performance of magnetohydrodynamic channels with the goal of maximizing the electric power output from the plasma through varying one or more of a multitude of geometric and operational parameters.

# Nomenclature

| | |
|---|---|
| Plasma | A gaseous conductor (ionized gas) with a sufficient level of charged particles (ions and free electrons) to allow the passage of electric current and the response to magnetic fields upon motion. |
| Partially-ionized plasma | A class of plasma where only a fraction of the molecules are ionized. Alternative terms include *weakly-ionized plasma*, and *low-density plasma*. Combustion-gas plasma, which is our interest, is an example. Thus, we focus here on this class of plasma. |
| Fully-ionized plasma | A class of plasma where the gases are completely ionized. Examples include solar winds, stellar interiors (e.g., the core of the Sun), and fusion plasma. |
| Equilibrium plasma | A class of plasma where the temperature of the free electrons is the same as the temperature of the ions. Thus the plasma gas is characterized by a single common temperature value. Due to their smaller mass, electrons are much more mobile than ions (the other current carrier particle in plasma). Thus, the energy of the ohmic heating (Joule heating) of the plasma gas is given to the electrons, which are normally maintained at essentially the same temperature as the ions and neutral particles as a result of the high collision frequencies and energy transfer per collision. Therefore, the free electrons and heavy gas particles (ions and neutrals) are in thermal equilibrium. An alternative term for this plasma class is *thermal plasma*, and *quasi-equilibrium plasma*. Examples include combustion-gas plasma [5] solar plasma, and arc discharge (electrical breakdown of a normally-nonconductive gas). We here focus on this class of plasma. |
| Non-Equilibrium plasma | A class of plasma where the temperature of the free electrons is higher than the temperature of the ions and neutrals. This situation occurs if ionizing current densities flowing through the plasma are sufficiently high, then the electron temperature will be significantly higher than that of the main body of the plasma gas. This is most easily attained in a monatomic gas with a large atom/electron mass ratio (typically argon), because in the absence of molecular vibrational excitation and the large disparity between the electron mass and heavy-particle mass, only a small fraction of the energy difference is exchanged on collision [39]. An alternative term for this plasma class is *non-thermal plasma*. Examples include aurora borealis, glow discharge (a plasma formed by the passage of electric current through a low-pressure gas), corona discharge (electrical discharge brought on by the ionization of a fluid surrounding a conductor that is electrically charged) [40]. |
| Faraday field | This is an induced electric field (and current density) due to the motion of plasma gas while subject to an applied magnetic field. |
| Hall field | This is another secondary induced electric field (and current density) due to the motion of the electrons (as a consequence of the primary Faraday current) as conductors while subject to the same applied magnetic field that induced the primary Faraday field. |
| Hall effect | The Hall effect describes the presence of the Hall field, leading to the inclination of the current density vector such that it is no longer collinear with the electrostatic field vector within the plasma. |
| $\vec{B}$ | Magnetic-field flux density (or simply the magnetic field, also called magnetic induction); tesla (T) = Wb/m$^2$ = V·s/m$^2$ = kg/s$^2$·A |
| $e$ | Elementary charge (proton charge or magnitude of electron charge); 1.602×10$^{-19}$ C [41] |
| $\vec{E_0}$ | Electrostatic field; V/m = N/C = kg·m/A·s$^3$ |
| $IE'$ | Ionization energy for an alkaline species (in eV, electron volts)<br>Note: 1 joule is equivalent to $\frac{1}{e\,=\,1.602\,\times\,10^{-19}} = 6.2415 \times 10^{18}$ eV [42] |
| $\vec{J}$ | Electric-current density (or simply the current density); A/m$^2$ |

The current density is calculated here as product of the electron charge, electron number density, and electron drift velocity [28]

$$\vec{J} = -\,e\,n_e \vec{v_d} \qquad (51)$$

The minus sign is a consequence of the negative charge of the electron. It is worth mentioning that the general form of Eq. (51) is

$$\vec{J} = e\,(n_{ion}\vec{u_{ion}} - n_e\vec{u_e}) \qquad (52)$$

where $n_{ion}$ and $\overrightarrow{u_{ion}}$ are the number density and absolute velocity of the ions, respectively as the other charge-carrying particle in the plasma, with a positive charge. However, since the ions are much heavier than electrons, their velocity is taken to be same as the bulk gas velocity; i.e., $\overrightarrow{u_{ion}} = \vec{u}$. Using a velocity relation (which is explained at a later part in the current section) that $\overrightarrow{u_e} = \vec{u} + \overrightarrow{v_d}$, the above equation can be written as [43, 44]

$$\vec{J} = e\,(n_{ion}\overrightarrow{u_{ion}} - n_e\,\vec{u} - n_e\,\overrightarrow{v_d}) \qquad (53)$$

Because the plasma is macroscopically neutral [3], the number of electrons is equal to the number of ions, leading to $n_{ion} = n_e$. These two assumptions lead to Eq. (51).

| | |
|---|---|
| $k_B$ | Boltzmann constant; $1.3806 \times 10^{-23}$ J/K [45] |
| $k_B'$ | Modified (expressed in eV/K) Boltzmann constant; $8.6173 \times 10^{-5}$ eV/K [46] |
| $h$ | Planck constant; $6.6261 \times 10^{-34}$ J·s [47] |
| $m_e$ | Electron mass; $9.109 \times 10^{-31}$ kg [48] |
| $n_e$ | Number density of electrons (average number of electrons in a unit volume); 1/m$^3$ |
| $\vec{u}$ | Absolute (that is being relative to laboratory/fixed frame) velocity vector of the plasma; m/s |
| $\overrightarrow{u_e}$ | Absolute (that is being relative to laboratory/fixed frame) velocity vector of the electrons; m/s |
| $\overrightarrow{v_d}$ | Drift velocity of the electrons (relative to the plasma medium); m/s |

In vacuum and in the absence of an electric field, the electron possesses a random motion with a zero average. In the presence of an electric field affecting an electron in vacuum, a net force acts on the electron accelerating it in the same direction of the electric field (e.g., toward the more-positive electrode in case of an electrostatic field). In the presence of an electric field in a material, the accelerated electron exhibits frequent collisions, leading to a finite average terminal velocity, called the *drift velocity*, which is used in the definition of the electron mobility. The drift velocity refers to an average velocity because actually the electron does not travel in a straight line, but its path is erratic (with curved segments between two collisions if an applied magnetic field is also present). The drift velocity is normally much smaller than the random velocity of the electrons, whose RMS value is $c_{e,rms} = \sqrt{\frac{3\,k_B\,T}{m_e}} = 6213\,\sqrt{T}$, where $c_{e,rms}$ is in m/s and $T$ is in kelvins.

From the principles of kinematics [49], we find that

$$\overrightarrow{u_e} = \vec{u} + \overrightarrow{v_d} \qquad (54)$$

| | |
|---|---|
| $t$ | Time |
| $x, y, z$ | Spatial rectangular coordinates; m |
| $\beta$ | Electron's Hall parameter (or simply Hall parameter); no unit |

In ionized gases (plasma) subject to a magnetic field, $\beta$ is the ratio of the electron gyrofrequency (the angular frequency of the electron circular motion in the plane perpendicular to an applied magnetic field, also called the cyclotron frequency) $\omega_e = \frac{e\,B}{m_e}$ (in rad/s), to the mean electron collision frequency $\nu_e$ (in Hz). It is also the ratio of the electron mean-free-path $\lambda_e$ to the gyroradius (the radius of the circular motion of the electron $r_e$, also called the Larmor radius) [50]. This Hall parameter is also the product of the electron mobility $\mu_e$ and the magnitude of the applied magnetic field $B$.

Moreover, for the simple configuration of axially-moving plasma moving in a horizontal constant-area channel with top anode (low-voltage electrode) and bottom cathode (high-voltage electrode) and applied magnetic field in the binormal direction (perpendicular to both the axial plasma motion and the vertical electrode-normal direction), then the induced electric current density vector $\vec{J}$ within the plasma will be vertically down if the Hall parameter is zero (practically very small) because of having a zero component in the axial direction. However, in general this current density will be inclined with an angle (Hall angle) $\theta_H$ from the vertical due to the presence of an axial component. This angle is the arctan of the Hall parameter (Figure 3). This feature is mathematically explained in detail in Appendix C.

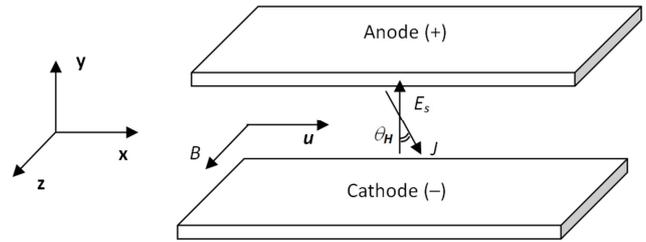

*Figure 3. Sketch depicting the Hall angle.*

Finally, the Hall parameter is also equal to the product of the gyrofrequency and the mean electron collision time $\left(\tau_e = \frac{1}{\nu_e}\right)$.

With this, the Hall parameter has multiple equivalent mathematical expressions,

$$\beta = \frac{\omega_e}{\nu_e} = \omega_e\,\tau_e = \frac{e\,B}{m_e\nu_e} = \frac{\lambda_e}{r_e} = \mu_e\,B = \tan(\theta_H) \qquad (55)$$

$\varepsilon$: Electric permittivity; farad/m (F/m) = (C/V)/m = A$^2$·s$^4$/kg·m$^3$

It is a medium property measuring the capacity of that medium to permit electric field lines. More specifically, it is the ratio of the generated electric displacement field (in C/m$^2$), manifested in slight separation of the positive and negative charges – atomic nuclei and their electrons – thus causing a local electric dipole moment, in response to an electric field (in V/m).

The electric permittivity of vacuum $\varepsilon_0$ (also called the permittivity of free space, and the electric constant) is a universal constant having the approximate value of ($8.854 \times 10^{-12}$ F/m [51]).

$\mu_e$: Electron mobility (or simply the mobility); (m/s)/(V/m) = m$^2$/V·s = T$^{-1}$

It is a material property measuring how quickly an electron moves through that material when pulled by an electric field. More specifically, it is the ratio of the average electron speed, the drift velocity, (in m/s) in response to an applied electric field (in V/m). The electron mobility is related to the magnitude of electron charge $e$, electron mass $m_e$, and mean electron collision frequency $\nu_e$ according to the following expression [52]:

$$\mu_e = \frac{e}{m_e \nu_e} \qquad (56)$$

$\mu_m$ : Magnetic permeability (or simply the permeability); henry/m (H/m) = (Wb/A)/m = kg·m/A$^2$·s$^2$

It is a medium property measuring the capability of that medium to be magnetized. More specifically, it is the ratio of the generated magnetic field (in Wb/m$^2$) in response to a driving magnetizing field (in A/m) produced by electric current flow in a coil of wire. The magnetic permeability of vacuum $\mu_0$ (also called the permeability of free space, and the magnetic constant) is a universal constant having the exact value of ($4\pi \times 10^{-7}$ H/m [53]).

The non-zero values of vacuum permittivity and vacuum permeability are consistent with the finiteness of the speed of light in vacuum $c_0$, where: $c_0 \sqrt{\varepsilon_0 \mu_0} = 1$. These non-zero values are attributed to the magnetization and the polarization of continuously appearing and disappearing fermion pairs, and the vacuum is hypothetically considered a medium filled with continuously appearing and disappearing charged fermion pairs [54].

$\rho_{ch}$ : Net electric-charge density due to ions and electrons (or simply the charge density); C/m$^3$. For plasma, $\rho_{ch} = 0$ because the plasma is globally neutral.

$\sigma$ : Plasma electric conductivity (or simply the conductivity); (A/m$^2$)/(V/m) = A/V·m = siemens/m (S/m) = 1/ohm·m (1/Ω·m) = A$^2$·s$^3$/kg-m$^3$

It is a material property measuring the capability of that material to conduct electricity. More specifically, it is the ratio of the generated current density in the material (in A/m) in response to a driving electric field (in V/m). The plasma electric conductivity is the product of the magnitude of electron charge $e$, electron number density $n_e$, and electron mobility $\mu_e$,

$$\sigma = e\, n_e \mu_e \qquad (57)$$

$\nu_e$ : Mean frequency of the electron collision with other particles (neutrals, ions, electrons); 1/s = Hz

## Acknowledgements

The author deeply appreciates the valuable cooperation with Ernest David Huckaby, Charles Rigel Woodside, and Geo Richards (National Energy Technology Laborator of the U.S. Department of Energy).

## Appendices

### Appendix A: Low Magnetic Reynolds Number

The behavior of the magnetic field depends largely on the level of electric conductivity of the medium. More precisely, it depends on a dimensionless number called the *magnetic Reynolds number* (Re$_m$). When Re$_m \gg 1$, the magnetic field lines act like elastic bands frozen into the conducting medium [18]. The velocity of the plasma induce a secondary magnetic field $\vec{b}$, to be added to the applied magnetic field $\vec{B}$. This flow-dependent magnetic field then affects the plasma velocity through the source term in the plasma momentum equation. Thus, there is a two-way coupling between the magnetic field and the velocity field. On the other hand, when Re$_m \ll 1$ the plasma velocity has a negligible effect on the magnetic field, whose profile is safely fixed. The coupling between the magnetic field and the plasma motion is thus of a one-way pattern, and the induced current densities in the plasma practically do not distort the applied magnetic field [25]. This low-Re$_m$ condition is also called inductionless approximation [55]. The Re$_m$ is defined as:

$$\text{Re}_m = \mu_m^* \sigma^* u^* L^* \qquad (A\text{-}1)$$

where $\mu_m^*$ is a reference magnetic permeability, $\sigma^*$ is a reference electric conductivity, $u^*$ is a reference plasma speed, and $L^*$ is a reference length relevant to the geometric scales.

Noting that $(\sigma^* u^*)^{-1}$ has a dimension of kinematic diffusivity (i.e., m$^2$/s), Eq. (A-1) can be re-written as

$$\text{Re}_m = \frac{u^* L^*}{\nu_m^*} \qquad (A\text{-}2)$$

where $\nu_m^*$ is a reference magnetic diffusivity. The above equation resembles the classical definition of the dimensionless Reynolds number used in fluid mechanics as a normalized flow velocity to classify the flow as either laminar or turbulent [56]. The strong similarity between the two dimensionless numbers justifies their similar names, although the Re$_m$ is not in fact devised by Osborne Reynolds after whom the fluid-mechanics dimensionless number is named. A small Re$_m$ is typical in MHD channels [25, 50, 57]. For example, hot air at 3273 K with 2% potassium and a reference length of 1 m and a velocity of 800 m/s has a Re$_m$ of 0.1 [28]. This assumption fails in astrophysics applications (solar and space plasma) because of the enormous reference lengths encountered there [58]. It also fails in fusion plasma (which is fully-ionized [43] with extremely large temperatures possibly exceeding $10^8$ K [59]) because of the large electric conductivity.

### Appendix B: Derivation of Generalized Ohm's Law

The forces acting on an electron with a charge of ($-e$) are [44, 60]
1. Momentum-exchange collision force [43]: $-m_e \nu_{eH} (\overrightarrow{u_e} - \overrightarrow{u_H}) = -m_e \nu_{eH} \overrightarrow{v_d}$
2. Electrostatic force (Coulomb force): $-e \vec{E_s}$
3. Magnetic-field force due to its flow-wise velocity with the plasma: $-e\, \vec{u} \times \vec{B}$
4. Magnetic-field force due to its drift (relative to plasma): $-e\, \overrightarrow{v_d} \times \vec{B}$

For the first force, $\nu_{eH}$ is the mean frequency of electron-heavy particle collision. The heavy particle velocity is taken to be equal to the plasma-gas velocity ($\overrightarrow{u_H} = \vec{u}$), given the relatively large mass of these heavy particles compared to the electrons. The last two forces combine together as a magnetic-field force due to the absolute electron velocity: $-e\,\overrightarrow{u_e}\times\vec{B}$

Applying the dynamic force balance equation with safely neglected inertial term (due to the small mass of the electron [58]), we obtain:

$$m_e\,\nu_{eH}\,\overrightarrow{v_d} = -e\left(\overrightarrow{E_s}+\vec{u}\times\vec{B}+\overrightarrow{v_d}\times\vec{B}\right) \quad (B\text{-}1)$$

Dividing the above equation by $-m_e\,\nu_{eH}$, and expanding the terms in the RHS, we obtain

$$\overrightarrow{v_d} = -\frac{e}{m_e\,\nu_{eH}}\overrightarrow{E_s} - \frac{e}{m_e\,\nu_{eH}}\vec{u}\times\vec{B} - \frac{e}{m_e\,\nu_{eH}}\overrightarrow{v_d}\times\vec{B} \quad (B\text{-}2)$$

Although the above equation is sufficiently detailed, it should be expressed in terms of the plasma macroscopic properties $(\sigma,\mu_e)$ and current density $(\vec{J})$. To do this, we utilize that

$$\mu_e = \frac{e}{m_e\,\nu_{eH}} \quad (B\text{-}3)$$

Now, Eq. (B-2) becomes

$$\overrightarrow{v_d} = -\mu_e\,\overrightarrow{E_s} - \mu_e\vec{u}\times\vec{B} - \mu_e\overrightarrow{v_d}\times\vec{B} \quad (B\text{-}4)$$

Multiplying the last equation by $(-e\,n_e)$, we obtain

$$(-e\,n_e\overrightarrow{v_d}) = \langle e\,n_e\,\mu_e\rangle\,\overrightarrow{E_s} + \langle e\,n_e\,\mu_e\rangle\,\vec{u}\times\vec{B} - \mu_e(-e\,n_e\overrightarrow{v_d})\times\vec{B} \quad (B\text{-}5)$$

Utilizing that

$$\vec{J} = (-e\,n_e\,\overrightarrow{v_d}) \quad (B\text{-}6)$$

for the LHS term and the last RHS term, while utilizing

$$\sigma = \langle e\,n_e\,\mu_e\rangle \quad (B\text{-}7)$$

for the first and second RHS terms, leading to

$$\vec{J} = \sigma\,\overrightarrow{E_s} + \sigma\,\vec{u}\times\vec{B} - \mu_e\,\vec{J}\times\vec{B} \quad (B\text{-}8)$$

which is the generalized Ohm's law given in Eq. (8).

### Appendix C: Effective-Conductivity Tensor (and Hall Angle)

We showed that the generalized Ohm's law can be expressed as

$$\vec{J} = \overrightarrow{\overrightarrow{\sigma_{eff}}}\cdot\left(\overrightarrow{E_s}+\vec{u}\times\vec{B}\right)$$

We here aim at giving the analytical expressions for the elements of $\overrightarrow{\overrightarrow{\sigma_{eff}}} = \sigma\,\overrightarrow{\overrightarrow{A^{-1}}}$, in terms of the electron mobility $\mu_e$ and the Cartesian components of the applied magnetic field vector $\vec{B} = [B_x,B_y,B_z]^T$.

The coefficient matrix $\vec{\vec{A}}$ has the following elements:

$$\vec{\vec{A}} = \begin{bmatrix} 1 & \mu_e B_z & -\mu_e B_y \\ -\mu_e B_z & 1 & \mu_e B_x \\ \mu_e B_y & -\mu_e B_x & 1 \end{bmatrix} \quad (C\text{-}1)$$

To simplify the expression for $\overrightarrow{\overrightarrow{A^{-1}}}$, let $x,y,z$ denote $\mu_e B_x,\mu_e B_y,\mu_e B_z$, respectively. Then

$$\vec{\vec{A}} = \begin{bmatrix} 1 & z & -y \\ -z & 1 & x \\ y & -x & 1 \end{bmatrix} \quad (C\text{-}2)$$

We found that $\overrightarrow{\overrightarrow{\sigma_{eff}}} = \sigma\,\overrightarrow{\overrightarrow{A^{-1}}}$ has the following analytical expression:

$$\overrightarrow{\overrightarrow{\sigma_{eff}}} = \frac{\sigma}{x^2+y^2+z^2+1}\begin{bmatrix} x^2+1 & x\,y-z & x\,z+y \\ x\,y+z & y^2+1 & y\,z-x \\ x\,z-y & y\,z+x & z^2+1 \end{bmatrix} \quad (C\text{-}3)$$

We point that the elements of both $\vec{\vec{A}}$ and $\overrightarrow{\overrightarrow{A^{-1}}}$ are dimensionless. In the special, but not trivial, case of having the magnetic field acting purely in the lateral direction, $\vec{B} = [0,0,B_z]^T$, Eq. (C-3) reduces to

$$\overrightarrow{\overrightarrow{\sigma_{eff}}}\bigg|_{Bz} = \frac{\sigma}{z^2+1}\begin{bmatrix} 1 & -z & 0 \\ z & 1 & 0 \\ 0 & 0 & z^2+1 \end{bmatrix} \quad (C\text{-}4)$$

and in this case, $B_z = B = \|\vec{B}\|$, thus $z \equiv \mu_e B_z$ is $\mu_e B$, which is the Hall parameter $\beta$. Thus, the above expression takes the form

$$\overrightarrow{\overrightarrow{\sigma_{eff}}}\bigg|_{Bz} = \sigma\begin{bmatrix} 1 & -\frac{\beta}{\beta^2+1} & 0 \\ \frac{\beta}{\beta^2+1} & 1 & 0 \\ 0 & 0 & 1 \end{bmatrix} \quad (C\text{-}5)$$

which decouples the x- and y- components of $\vec{J}$ from its z-component. Applying the above expression in the generalized Ohm's law, Eq. (15), we obtain the following scalar equations for the Cartesian components of the current density vector:

$$J_x = \frac{\sigma}{\beta^2+1}\left[(E_{s,x}+v\,B_z)-\beta\,(E_{s,y}-u\,B_z)\right]$$

$$J_y = \frac{\sigma}{\beta^2+1}\left[(E_{s,y}-u\,B_z)+\beta\,(E_{s,x}+v\,B_z)\right]$$

$$J_z = \sigma\,E_{s,z} \quad (C\text{-}6)$$

where $E_{s,x}, E_{s,y}$ and $E_{s,z}$ are the x-, y-, and z- components of the electrostatic field, respectively; and $u$ and $v$ are the x- and y- components of the plasma velocity vector, respectively. Again, $(J_z)$ is decoupled from $(J_x, J_y)$. It is worth noting that the z- component (aligned with the magnetic field) of the plasma velocity now has no influence on the electric aspects of the problem.

Proceeding further with Eq. (C-6); in the simplified case of one-dimensional plasma (i.e., $\vec{u} = [u,0,0]^T$), and one-dimensional electrostatic field (i.e., $\overrightarrow{E_s} = [0,E_{s,y},0]^T$), the equations reduce further to

$$J_x = -\sigma \frac{\beta}{\beta^2 + 1} (E_{s,y} - u B_z) = \beta |J_y|$$

$$|J_y| = -J_y = \sigma \frac{1}{\beta^2 + 1} (u B_z - E_{s,y})$$

$$J_z = 0 \qquad (C\text{-}7)$$

This model case demonstrates the unfavorable effects of the Hall parameter, where it (1) causes a parasite Hall current density $J_x$, and (2) reduces the magnitude of the useful Faraday current density $|J_y|$. It also explains the presence of the Hall angle $\theta_H$ for the current density vector $\vec{J}$ (measured from the vertical y-axis) in this case, as given in Figure 3, with $\theta_H = \tan^{-1}(\beta)$.

*Appendix D: Analytical Fits for Collision Cross-Sections*

In the following expressions, a prime in a symbol emphasizes that its unit is not according to the SI system of units. The electron-molecule collision cross-section for each gaseous species ($Q'_{species}$ in cm$^2$) is expressed as a function of the electron energy $u'$ (in eV), calculated in turn as a function of the absolute temperature $T$ (in kelvins) as follows:

$$u' = 1.5 \, k'_B \, T \qquad (D\text{-}1)$$

For all species except argon, the Frost expression calculates the product $(Q'_{species} \, C'_e)$ in cm$^3$/s, where

$$C'_e = 100 \, C_e = 100 \, C_{ce} \sqrt{T} \qquad (D\text{-}2)$$

is the mean magnitude of electron random velocity, expressed in cm/s.

1. For argon (Ar), we derived an eighth-degree polynomial fit given below

$$Q'_{Ar} = 10^{-16} \exp(a_0 \, \xi + a_1 \, \xi + a_2 \, \xi^2 + a_3 \, \xi^3 + a_4 \, \xi^4 + a_5 \, \xi^5 + a_6 \, \xi^6 + a_7 \, \xi^7 + a_8 \, \xi^8)$$

where $\xi = \ln(u), a_0 = -0.04489, a_1 = 2.685, a_2 = 10.95, a_3 = 34.76,$

$$a_4 = 48.31, a_5 = 34.44, a_6 = 13.33, a_7 = 2.663, a_8 = 0.2148 \qquad (D\text{-}3)$$

2. For carbon monoxide (CO)

$$Q'_{CO} \, C'_{ce} = 9.1 \times 10^{-8} \, u' \qquad (D\text{-}4)$$

3. For carbon dioxide (CO$_2$)

$$Q'_{CO2} \, C'_{ce} = 10^{-8} \left(\frac{1.7}{\sqrt{u'}} + 2.1 \sqrt{u'}\right) \qquad (D\text{-}5)$$

4. For cesium (Cs)

$$Q'_{Cs} \, C'_{ce} = 10^{-8} (160) \qquad (D\text{-}6)$$

5. For hydrogen atom (H)

$$Q'_H \, C'_{ce} = 10^{-8} \left(42 \sqrt{u'} - 14 \, u'\right) \qquad (D\text{-}7)$$

6. For hydrogen molecule (H$_2$)

$$Q'_{H2} \, C'_{ce} = 10^{-8} \left(4.5 \sqrt{u'} + 6.2 \, u'\right) \qquad (D\text{-}8)$$

7. For water vapor (H$_2$O)

$$Q'_{H2O} \, C'_{ce} = 10^{-8} \left(\frac{10}{\sqrt{u'}}\right) \qquad (D\text{-}9)$$

8. For helium (He)

$$Q'_{He} \, C'_{ce} = 10^{-8} \left(3.14 \sqrt{u'}\right) \qquad (D\text{-}10)$$

9. For potassium (K)

$$Q'_K \, C'_{ce} = 10^{-8} (160) \qquad (D\text{-}11)$$

10. For nitrogen molecule (N$_2$)

$$Q'_{N2} \, C'_{ce} = 10^{-8} (12 \, u') \qquad (D\text{-}12)$$

11. For neon (Ne)

$$Q'_{Ne} \, C'_{ce} = 10^{-8} (1.15 \, u') \qquad (D\text{-}13)$$

12. For oxygen atom (O)

$$Q'_O \, C'_{ce} = 10^{-8} \left(5.5 \sqrt{u'}\right) \qquad (D\text{-}14)$$

13. For oxygen molecule (O$_2$)

$$Q'_{O2} \, C'_{ce} = 10^{-8} \left(2.75 \sqrt{u'}\right) \qquad (D\text{-}15)$$

14. For hydroxide group (OH)

$$Q'_{OH} \, C'_{ce} = 10^{-8} \left(\frac{8.1}{\sqrt{u'}}\right) \qquad (D\text{-}16)$$